\documentclass[pdflatex,sn-mathphys-num]{sn-jnl}
\usepackage{graphicx}%
\usepackage{multirow}%
\usepackage{amsmath,amssymb,amsfonts}%
\usepackage{amsthm}%
\usepackage{mathrsfs}%
\usepackage[title]{appendix}%
\usepackage{xcolor}%
\usepackage{textcomp}%
\usepackage{manyfoot}%
\usepackage{booktabs}%
\usepackage{algorithm}%
\usepackage{algorithmicx}%
\usepackage{algpseudocode}%
\usepackage{listings}%
\DeclareMathOperator{\sgn}{sgn}
\DeclareMathOperator{\intfun}{int}
\DeclareMathOperator{\maxfun}{max}

\begin{document}

\title{Casimir effect in Josephson junctions}

\author{\fnm{Alex} \sur{Levchenko}}

\affil{\orgdiv{Department of Physics}, \orgname{University of Wisconsin-Madison}, \orgaddress{\city{Madison}, \postcode{53706}, \state{WI}, \country{USA}}}

\abstract{In a Josephson junction, the supercurrent is determined by both the discrete sub-gap part of the spectrum due to Andreev bound states and the continuous part of the spectrum from energy states outside the superconducting gap. We consider the cohesive force exerted on a junction, which is thermodynamically conjugated to the superflow, and comment on its connection to the Casimir effect in quantum electrodynamics. In contrast to the supercurrent, it is shown that in ballistic short junctions, the force is predominantly contributed by the continuum. Its magnitude is universally defined by the energy gap and coherence length of the superconductor per spin-dependent transverse mode. This force scales non-analytically with the junction length and is periodic with the superconducting phase. For long ballistic junctions, the force results from the interplay of oscillatory contributions originating from both bound states and the continuum. The resulting asymptotic limit for the force is established, including the correction terms. Thermal and impurity effects on the force are briefly discussed.  
\footnote{Submitted to the special issue of JLTP for publication as part of the Memorial Collection in honor of Alexander Andreev.} \newline

\centering{Dated: October 10, 2024}}


\maketitle

\section{Introduction} 

In a Josephson junction, the flow of a supercurrent is directly related to the thermodynamically conjugated cohesive force \cite{Krive:2004a}. Indeed, given the knowledge of the grand canonical potential 
$\Omega$ for the system, both the current $I$ and the force $F$ can be found from the following relations:
\begin{equation}\label{eq:I-F}
I=\frac{2e}{\hbar}\frac{\partial\Omega}{\partial\phi}, \quad F=-\frac{\partial\Omega}{\partial L},
\end{equation}
where $\phi$ is the superconducting phase difference applied across the junction, and $L$ is the length of the junction between superconducting reservoirs. Then, considering the differential form of $\Omega$, one can infer the corresponding Maxwell relation 
\begin{equation}\label{eq:Maxwell}
\frac{\partial F}{\partial\phi}=-\frac{\hbar}{2e}\frac{\partial I}{\partial L}.
\end{equation}
Therefore, knowledge of the current-phase relation yields the force and vice versa. The Josephson current has been meticulously studied over the years in various systems and superconducting devices (see reviews \cite{GKI:2004,Buzdin:2005,Beenakker:2008,Nadeem:2023} and references therein), in contrast, the focus on the force has received comparatively little attention. 

In complete analogy to the supercurrent, the physical origin of this force can be traced back to the quantization of electron energy levels in the junction due to coherent Andreev reflections \cite{Andreev:1964} from superconducting boundaries, which lead to the formation of Andreev bound states (ABS), also known as Andreev-Kulik levels \cite{Kulik:1970}. Interestingly, this force also admits a quantum field-theoretic interpretation in relation to the Casimir effect \cite{Casimir:1948} in quantum electrodynamics.

Specifically, the Andreev boundary condition alters the energy of the Fermi sea of quasiparticles in the normal region of the Josephson junction. The shift in ground state energy induced by these boundary conditions defines the Casimir energy $\Omega_{\text{C}}$. For a massless field, dimensional considerations suggest a simple scaling of this energy with the geometrical size of the system. In one dimension, it is of the order of $\sim \hbar v/L$, where 
$v$ is the speed of light for a relativistic field or the Fermi velocity for electrons. By calculating this shift for the spectrum of Andreev-Kulik states one can determine the numerical prefactor in the Casimir energy estimate and establish its dependence on the superconducting phase. The result is \cite{Krive:2004b}
\begin{equation}
\Omega_{\text{C}}=g\pi\hbar\frac{v}{2L}\left[\left(\frac{\phi}{2\pi}\right)^2-\frac{1}{12}\right],\quad |\phi|<\pi,
\end{equation}
per ballistic transverse mode, where the factor $g$ counts spin and other degeneracies. From Eq. \eqref{eq:I-F} one then determines both the current and the force at zero temperature ($T=0$): 
\begin{equation}\label{eq:-I-F-sawtooth}
I=\frac{gev}{L}\left(\frac{\phi}{2\pi}\right),\quad F=\frac{\pi g\hbar v}{2L^2}\left[\left(\frac{\phi}{2\pi}\right)^2-\frac{1}{12}\right],\quad |\phi|<\pi. 
\end{equation}
The result for the current coincides with the well-known sawtooth current-phase-relation \cite{Ishii:1970,Bardeen:1972,Bratus:1973} characteristic of a ballistic superconductor-normal metal-superconductor (SNS) Josephson junction in the long-junction limit when $L>\xi$ exceeds the scale of superconducting coherence length $\xi=\hbar v/\Delta$, where $\Delta$ is the superconducting energy gap (see also Refs. \cite{Svidzinsky,Zagoskin}). The conjugated force is attractive for $\phi=0$, which is precisely the Casimir force between perfectly reflecting metallic mirrors as produced by vacuum fluctuations of a scalar one-dimensional field. The force becomes repulsive at $\phi=\pi$, which in the context of the electromagnetic Casimir effect corresponds to a different boundary condition imposed on the field, namely Neumann boundary condition on one mirror and Dirichlet on the other mirror, see Ref. \cite{Boyer:2003} and more extensive literature overview on this subject \cite{Plunien:1986,Mostepanenko}. 

It is worth noting that Eq. \eqref{eq:-I-F-sawtooth} is independent of the superconducting parameters $\Delta$ and $\xi$, therefore, for temperatures $T\ll\Delta$ all the thermodynamic properties of the long ballistic junction are essentially universal and material independent. This observation, however, is correct only concerning the contribution of ABS to the thermodynamic properties in a particular limit. It will be shown below that the continuous spectrum (CS) also contributes, and the resulting expression can be found as a controlled expansion in $\xi/L$.     

The situation is even more subtle in the opposite limit of a short junction, $L\ll\xi$, which has no electromagnetic analog. Indeed, for a point contact where $L/\xi\to0$, the spectrum of ABS becomes independent of $L$. In this case, the Josephson current reaches its universal limit \cite{BvH:1991,Furusaki:1991,Beenakker:1991}: the critical current increases stepwise as a function of the constriction width, with a quantized step height of $e\Delta/\hbar$ at $T=0$, independent of the junction's properties. The total critical current is therefore determined by the total number of transverse channels $N_{\text{ch}}$, which counts the propagating modes at the Fermi level traversing the constriction. In this limit, the cohesive force vanishes $F\to0$. To establish its magnitude, one must investigate the corrections to the spectrum of ABS for a small but finite ratio between the junction length and coherence length, $L/\xi\ll1$. This analysis was carried out in Ref. \cite{Krive:2004a}, where it was shown that the cohesive elongation (or contraction) force exerted on a junction, which will be referred to as the Andreev-Casimir force hereafter, is of the order of $F\sim gN_{\text{ch}}(\Delta/\xi)|\sin\phi|$. However, at finite $L/\xi$, the continuum of states above the gap $\Delta$ also contributes to the supercurrent, and thus to the force. This is a well-established fact in the context of chaotic cavities \cite{Brouwer:1997}, diffusive junctions \cite{Levchenko:2006}, and quantum dots \cite{Fatemi:2024}, so it is not exclusive to the ballistic limit. This contribution was neglected in Ref. \cite{Krive:2004a}, which prompted a revision reported in the recent work Ref. \cite{Beenakker:2023}. The analysis, based on the scattering matrix formulation, revealed that the continuous spectrum provides the dominant contribution to the force, which is larger than that due to ABS by a logarithmic factor $\sim\ln(\xi/L)$. This result is confirmed by a different method in this work.

The objective of this paper is to reconsider the relative importance of the contributions to the Andreev-Casimir force from both the bound states and the continuum. 
From a practical standpoint, given the extensive knowledge and vast literature devoted to the computation of the Josephson current, it is more straightforward to calculate the current first and then derive the force using Maxwell's relation, Eq. \eqref{eq:Maxwell}. This approach forms the basis of the analysis presented here. This will allow us to review the existing results in the current literature and gain a deeper perspective on the problem across various limiting cases. To achieve this, in Sec. \ref{sec:Josephson} we will focus solely on the ballistic limit of the Josephson effect in both long ($L>\xi$) and short ($L<\xi$) junctions. In Sec. \ref{sec:Force} we compute asymptotic expressions for the corresponding quantum force. We also consider an impure SNS bridge with a single scatterer and establish the dependence of the force on the transparency of the junction. In diffusive (disordered) systems, the force is negligibly small as discussed in Sec. \ref{sec:Summary}. Finally, practical estimates are provided in Sec. \ref{sec:Summary}, along with a brief discussion of the experimental situation.

%#########################################################################
%#########################################################################
%#########################################################################
%#########################################################################
%#########################################################################

\section{Josephson current}\label{sec:Josephson}

In a generic Josephson junction the current-phase-relation contains all harmonics of the superconducting phase \cite{GKI:2004}
\begin{equation}
I=\sum^{\infty}_{n=1}[I_n\sin(n\phi)+I'_n\cos(n\phi)].
\end{equation}
The phase even harmonics $I'_n$ can be present only in the situation with broken time reversal symmetry. We exclude this situation in this work and set $I'_n\to0$. Retaining these terms leads to the so-called anomalous Josephson effect or equivalently $\phi_0$-junction \cite{Buzdin:2008}. The phase odd harmonics admit the following spectral representation (hereafter we use natural units $k_B=\hbar=1$ for simplicity)
\begin{equation}\label{eq:In}
I_n=-\frac{2G}{e}\int^{\frac{\pi}{2}}_{0}d\theta\cos\theta\int d\varepsilon \tanh(\varepsilon/2T)\mathcal{I}_n(\varepsilon,\theta), 
\end{equation}  
here $\theta$ measures the incidence angle of the electron/hole trajectory with respect to the normal of SN boundary while $G=\frac{e^2}{2\pi}gN_{\text{ch}}$ defines the normal state conductance of the junction. Here we envision a multichannel junction of width $W$ so that the number of channels is $N_{\text{ch}}=k_FW/\pi$, where $k_F$ is the Fermi momentum. For example, such junctions can be realized in devices with graphene as a normal layer, e.g. see \cite{Heersche:2007,Andrei:2008,Coskun:2012,Finkelstein:2016,Mason:2016}. 

The spectral supercurrent harmonics take different values for subgap or above the gap ranges of energies, specifically \cite{GKI:2004,Beenakker:1992}: 
\begin{subequations}\label{eq:Jn}
\begin{equation}
\mathcal{I}_n=\mathcal{I}^{\text{ABS}}_n+\mathcal{I}^{\text{CS}}_n,
\end{equation}
where
\begin{equation}\label{eq:Jn-ABS}
\mathcal{I}^{\text{ABS}}_n(\varepsilon,\theta)=\sin\left(\frac{2n\varepsilon L}{v\cos\theta}-2n\alpha(\varepsilon)\right) \quad \text{for} \quad |\varepsilon|<\Delta, 
\end{equation}   
and 
\begin{equation}\label{eq:Jn-CS}
\mathcal{I}^{\text{CS}}_n(\varepsilon,\theta)=a^{2n}(\varepsilon)\sin\left(\frac{2n\varepsilon L}{v\cos\theta}\right) \quad \text{for} \quad |\varepsilon|>\Delta.
\end{equation}   
\end{subequations}
Here the angular phase shift $\alpha(\varepsilon)=\arccos(\varepsilon/\Delta)$ and amplitude $a(\varepsilon)=|\varepsilon|/\Delta-\sqrt{(\varepsilon/\Delta)^2-1}$ describe Andreev reflections. It should be noted that: (i) spectral currents are odd functions of the quasiparticle energy $\mathcal{I}_n(-\varepsilon,\theta)=-\mathcal{I}_n(\varepsilon,\theta)$; (ii) amplitudes of Andreev reflections for high-harmonics rapidly vanish for energy states deep into the continuum $a^{2n}\approx \exp(-2n\ln(\varepsilon/\Delta))$ for $\varepsilon\gg\Delta$ and $n>1$; (iii) in point contacts, $L\to0$, continuum does not contribute to the current, while the contribution of the sub-gap spectrum is dominated by the phase shift $\alpha(\varepsilon)$; (iv) the Andreev-Kulik spectrum is restored in the long junctions where phase shifts are essentially energy independent for states deep in the gap $\alpha\approx\pi/2$ for $\varepsilon\ll\Delta$. (v) The spectral currents are continuous function across the gap since $\alpha(\Delta)=0$ and $a(\Delta)=1$, so that $\mathcal{I}^{\text{ABS}}_n(\Delta,\theta)=\mathcal{I}^{\text{CS}}_n(\Delta,\theta)$.  

These observations provide all the necessary ingredients to compute the integral in Eq. \eqref{eq:In}, using the input from Eq. \eqref{eq:Jn}, to determine both the current and the force. We approach this calculation in two limiting cases. 

%#########################################################################
%#########################################################################
%#########################################################################
%#########################################################################
%#########################################################################
\section{Andreev-Casimir force}\label{sec:Force}

%############################ 
\subsection{Short junctions}
%############################ 

We begin with the consideration of the short junction. In the limit of the ballistic point contact, $L\to0$, the contribution to the current from the continuum vanishes, since $\mathcal{I}^{\text{CS}}_n\to0$ as is evident from Eq. \eqref{eq:Jn-CS}.  The contribution from the subgap states is finite but it is independent of $L$ in this limit as it follows from Eq. \eqref{eq:Jn-ABS}. Therefore, in order to determine the force, we need to trace corrections to the subgap spectrum at finite $L/\xi\ll1$ and retain the contribution from the continuum to the leading order. 

To this end, we proceed with Eq. \eqref{eq:Jn-ABS} and expand it at small length
\begin{equation}
\mathcal{I}^{\text{ABS}}_n\approx-\sin(2n\alpha(\varepsilon))+\frac{2n\varepsilon L}{v\cos\theta}\cos(2n\alpha(\varepsilon)).
\end{equation}
Inserting this expressions into Eq. \eqref{eq:In} for the harmonics of the Josephson current we obtain at $T=0$:
\begin{subequations}
\begin{equation}
I^{\text{ABS}}_n=\left(\frac{4G\Delta}{e}\right)\int^{1}_{0}\sin(2n\arccos(x))dx,
\end{equation}
\begin{equation}
\delta I^{\text{ABS}}_n=-\left(\frac{4\pi G\Delta^2}{eE_{\text{Th}}}\right)n\int^{1}_{0}x\cos(2n\arccos(x))dx,
\end{equation}
\end{subequations}
where we introduced the ballistic Thouless energy $E_{\text{Th}}=v/L$.

The first term $I^{\text{ABS}}_n$ corresponds to the unperturbed spectrum in the point contact limit. Indeed, by changing the integration variable $y=\arccos x$, the integral can be reduced to the product of two  
sine functions $\sin(2ny)\sin(y)$ over $y\in[0,\pi/2]$. This integral is elementary and leads to the full current as a sum of harmonics that can be evaluated to the well-known result \cite{KO:1978,BvH:1991}  
\begin{equation}
I^{\text{ABS}}=\frac{2G\Delta}{e}\sum^{\infty}_{n=1}(-1)^{n+1}\frac{4n\sin n\phi}{4n^2-1}=\frac{\pi G\Delta}{e}\sin\frac{\phi}{2}\sgn\left(\cos\frac{\phi}{2}\right). 
\end{equation}
The characteristic feature of this expression is a non-sinusoidal current-phase relation with discontinuity at $\phi=\pi$. At finite temperature, $\sgn(\cos(\phi/2))\to\tanh(\Delta\cos(\phi/2)/2T)$, this discontinuity is smeared. 
The current is proportional to the normal state conductance of the junction and independent of its length, which corresponds to the vanishing force in this particular approximation, $F\propto \partial_L I\to0$. 

The second term $\delta I^{\text{ABS}}_n$ gives the first order correction at finite length, thus contributes to the force. Integrating over $x$ 
\begin{equation}
\delta I^{\text{ABS}}=\frac{\pi G\Delta^2}{eE_{\text{Th}}}\sum_{n=1}^\infty\frac{(1+(-1)^n)n\sin n\phi}{n^2-1}, 
\end{equation}
summing over $n$, and using Maxwell's relation we obtain force in the form 
\begin{equation}\label{eq:F-ABS-short}
F^{\text{ABS}}=\frac{1}{4}gN_{\text{ch}}\frac{\Delta}{\xi}\sum_{n=1}^\infty\frac{(1+(-1)^n)\cos n\phi}{n^2-1}. 
\end{equation}

\begin{figure}[t!]
\centering
\includegraphics[width=0.475\linewidth]{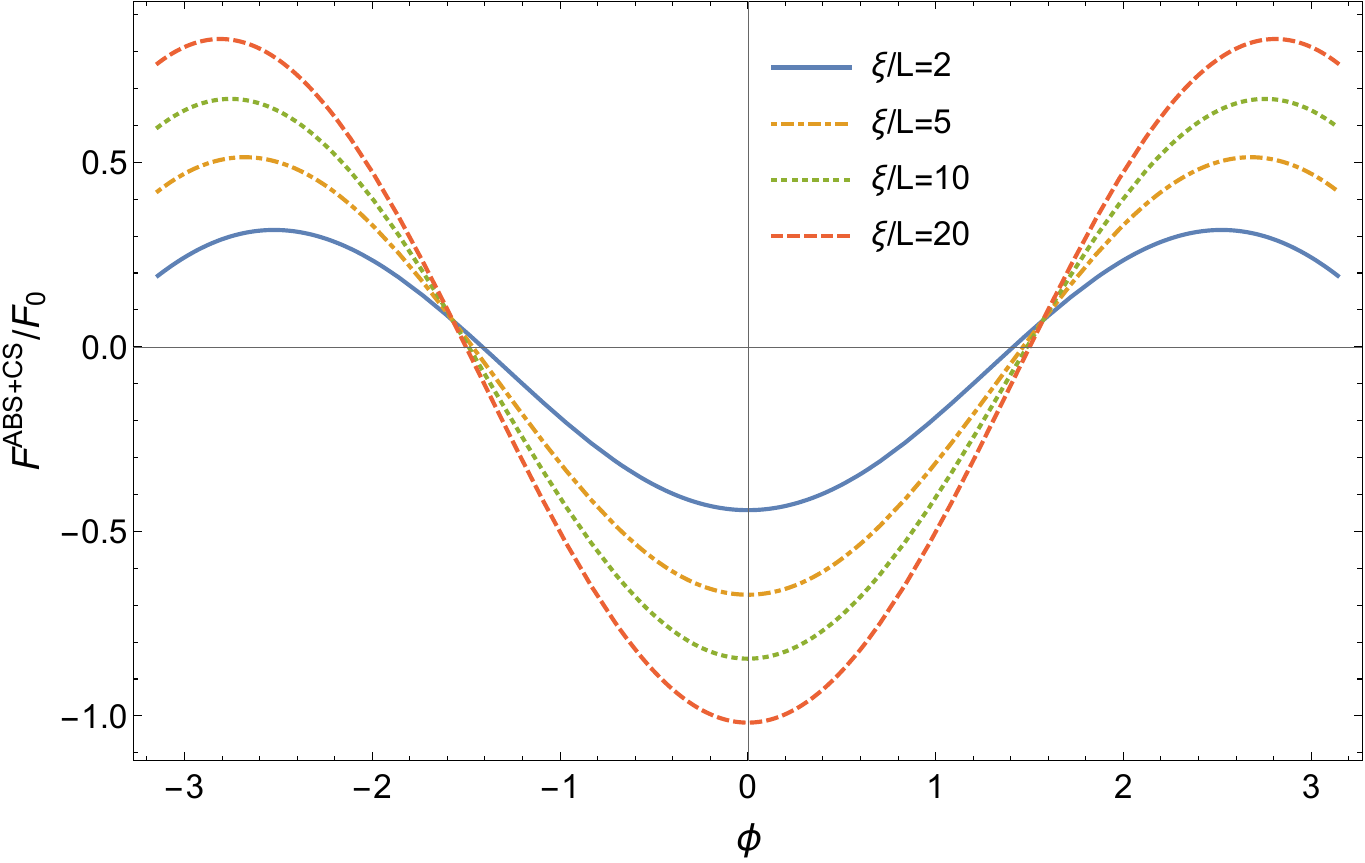}
\includegraphics[width=0.475\linewidth]{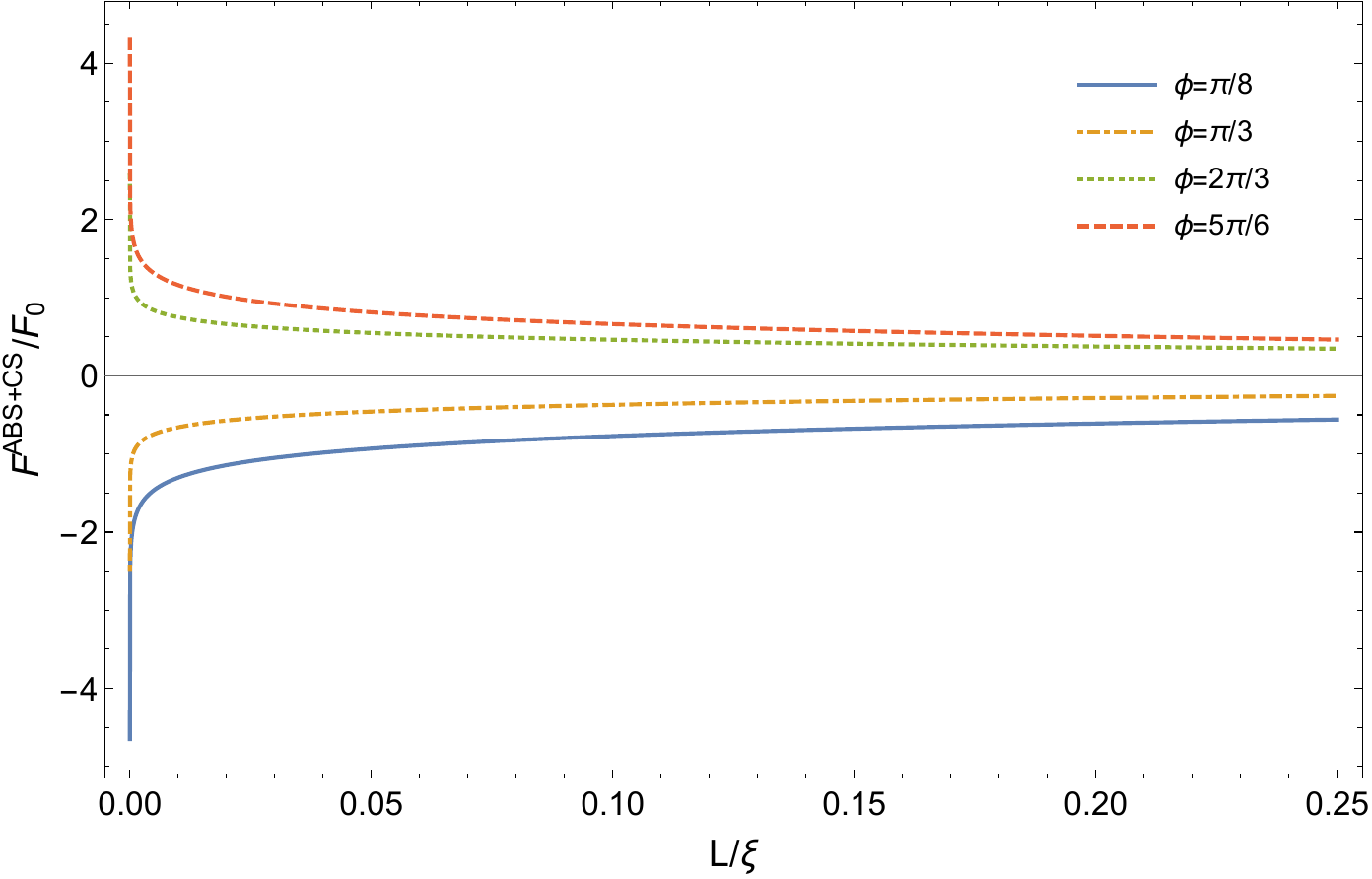}
\caption{The left panel shows the asymptotic total Andreev-Casimir force in the short-junction limit as a function of the superconducting phase, plotted for
$\xi/L=2,5,10,20$. The force is normalized to the unit of $F_0=gN_{\text{ch}}(\Delta/\xi)$. The right panel displays the same result plotted as a function of junction length for $\phi=\pi/8, \pi/3, 2\pi/3, 5\pi/6$. }\label{fig:F-short}
\end{figure}

In order to capture the leading contribution of the continuous states we use Eq. \eqref{eq:Jn-CS} where it is sufficient to approximate $\sin(2n\varepsilon L/v\cos\theta)\approx 2n\varepsilon L/v\cos\theta$ and take $a^{2n}(\varepsilon)\approx(\Delta/2\varepsilon)^{2n}$. Since integration expands over the energies $\varepsilon>\Delta$ it is evident that higher harmonics will be suppressed. Therefore, we retain only the first harmonic. Furthermore, the expansion of the sine function puts a bound on the upper limit of the integral as we have to require $\varepsilon\lesssim E_{\text{Th}}$. Putting everything together, we have for the first harmonic of the supercurrent 
\begin{equation}
I^{\text{CS}}_{n=1}\approx-\frac{4G}{e}\int^{\frac{\pi}{2}}_{0}d\cos\theta\int^{\sim \frac{E_{\text{Th}}}{\Delta}}_{1}\left(\frac{1}{2x}\right)^2\frac{2x\Delta^2dx}{E_{\text{Th}}\cos\theta}
\end{equation}   
Computing the remaining integral with the logarithmic accuracy, and restoring the force from Eq. \eqref{eq:Maxwell} we find to the leading order 
\begin{equation}\label{eq:F-CS-short}
F^{\text{CS}}\approx-\frac{1}{2\pi}gN_{\text{ch}}\frac{\Delta}{\xi}\ln\frac{E_\text{Th}}{\Delta}\cos\phi.
\end{equation} 
This result is larger than Eq. \eqref{eq:F-ABS-short} by a logarithmic factor, which confirms recent calculation \cite{Beenakker:2023} obtained via a different method based on the scattering matrix approach. 
To make an explicit connection, the numerical coefficient in above formula is computed for a strictly 1D channel. At finite temperature the logarithmic factor in Eq. \eqref{eq:F-CS-short} takes the form $\ln(E_{\text{Th}}/\maxfun\{\Delta,T\})$. 

The component of the force is negative for $\phi<\pi/2$, so it is a contraction force. It changes sign and crosses over to an elongation force for $\phi>\pi/2$. The total force, that is sum of $F^{\text{ABS}}$ and $F^{\text{CS}}$, with the correction terms up to $O(L/\xi)$, is plotted in Fig. \ref{fig:F-short} for several different ratios of $L/\xi$ and for different values of the superconducting phase $\phi$.  

%############################    
\subsection{Long junctions}
%############################    

For long junctions, $L>\xi$, focusing first on the Andreev-Kulik states deep inside the gap, the phase factor in Eq. \eqref{eq:Jn-ABS} can be approximated by a constant $\alpha(\varepsilon)\approx\pi/2$ for $\varepsilon\ll\Delta$.   
This step then gives for the harmonics in Eq. \eqref{eq:In}
\begin{equation}\label{eq:In-ABS-long}
I^{\text{ABS}}_n=(-1)^{n+1}\frac{2G}{e}\int^{\frac{\pi}{2}}_{0}d\theta\cos\theta\Im\int d\varepsilon \tanh(\varepsilon/2T)\exp(2in\varepsilon L/v\cos\theta)
\end{equation}
where we used the oddness of $\mathcal{I}_n$ in energy to bring sine into the exponential function. Formally speaking, the energy integral in this expression is bounded to $\in[-\Delta,\Delta]$. 
However, in the limit when $\Delta\gg E_{\text{Th}}$, one may stretch the domain of integration to energies above the gap $\varepsilon>\Delta$. The implications of this approximation will be discussed below. 
As the next step we use the tabulated integral 
\begin{equation}
\int\limits^{+\infty}_{-\infty} d\varepsilon \tanh(\varepsilon/2T)\exp(2in\varepsilon L/v\cos\theta)=\frac{2\pi iT}{\sinh\left(\frac{2\pi nLT}{v\cos\theta}\right)},
\end{equation}  
and obtain as a result for the supercurrent
\begin{equation}\label{eq:I-ABS-long}
I^{\text{ABS}}=\frac{2GE_{\text{Th}}}{e}\sum^{\infty}_{n=1}(-1)^{n+1}\left\langle\frac{\lambda_T\sin(n\phi)}{\sinh(n\lambda_T)}\right\rangle,\quad \lambda_T=\frac{2\pi T}{E_{\text{Th}}\cos\theta},
\end{equation} 
where angular brackets $\langle...\rangle$ denote angular average weighted with $\cos^2\theta$. It is clear that mainly trajectories with the normal incidence contribute to the angular integral over $\theta$, whereas the contribution of gliding trajectories along the contact is strongly suppressed. Eq. \eqref{eq:I-ABS-long} describes the thermally broadened sawtooth current-phase-relation. 
Indeed, setting $T\to0$, one recovers 
\begin{equation}\label{eq:I-sawtooth}
I^{\text{ABS}}=\frac{\pi GE_{\text{Th}}}{2e}\sum^{\infty}_{n=1}(-1)^{n+1}\frac{\sin n\phi}{n}=\frac{\pi GE_{\text{Th}}}{4e}[\phi], \quad [\phi]=\phi-\intfun(\phi/2\pi). 
\end{equation}
It is perhaps worth noting that evaluation of the Josephson current in this limit is equivalent to the calculation of the persistent current in a ring. 

Applying Eq. \eqref{eq:I-ABS-long} in Eq. \eqref{eq:Maxwell} to extract the corresponding force, one finds the following asymptotic (up to a numerical factor), 
\begin{equation}\label{eq:F-ABS-long}
F^{\text{ABS}}\simeq gN_{\text{ch}}\frac{\Delta}{\xi}\left\{\begin{array}{cc} \frac{E^2_{\text{Th}}}{\Delta^2}\left(\phi^2-\frac{\pi}{3}\right) & \quad\text{for}\quad T\ll E_{\text{Th}}\ll\Delta, \\ 
-\frac{T^2}{\Delta^2}e^{-2\pi T/E_{\text{Th}}}\cos\phi & \quad \text{for}\quad E_{\text{Th}}\ll T\ll\Delta,\end{array}\right.
\end{equation}  
which naturally complements the discussion presented in the introduction that for $T\to0$ force $F^{\text{ABS}}\propto1/L^2$ as in Eq. \eqref{eq:-I-F-sawtooth}. 

As the next step, we investigate the contribution coming from the continuum at energies $\varepsilon>\Delta$. We narrow this consideration only to $T=0$ limit as thermal smearing tends to suppress the effect. We thus have from Eqs. \eqref{eq:In} and \eqref{eq:Jn-CS}
\begin{equation}\label{eq:In-CS}
I^{\text{CS}}_n=-\frac{4G}{e}\int^{\frac{\pi}{2}}_{0}d\theta\cos\theta\int^{\infty}_{\Delta}d\varepsilon\, a^{2n}(\varepsilon)\sin\left(\frac{2n\varepsilon L}{v\cos\theta}\right).
\end{equation}
In order to proceed, we notice that the integral is a rapidly oscillating function. We change the order of integrations, $\theta\leftrightarrow\varepsilon$, and calculate the angular average by using the method of the stationary phase. For $f(\theta)=1/\cos\theta$ the stationary point is at $\theta=0$, namely the leading contribution to the current comes from the normal incidence, therefore, approximating $f\approx 1+\theta^2/2$ and taking $\cos\theta\to1$ in the prefactor we get   
\begin{equation}
I^{\text{CS}}_n\approx-\frac{2G\Delta}{e}\int^{\infty}_{1}dx a^{2n}(x)\Im\int^{+\infty}_{-\infty}\exp\left[\frac{2inx\Delta}{E_{\text{Th}}}(1+\theta^2/2)\right]d\theta
\end{equation} 
where we introduced the dimensionless variable $x=\varepsilon/\Delta$. In this expression, the $x$-integral is tabulated as the Fresnel integral from optics, which gives us, after taking its imaginary part  
\begin{equation}
I^{\text{CS}}_n\approx-\frac{2G\Delta}{e}\sqrt{\frac{\pi E_{\text{Th}}}{n\Delta}}
\int^{\infty}_{1}dx \frac{a^{2n}(x)}{\sqrt{x}}\sin\left(2nx\frac{\Delta}{E_{\text{Th}}}+\frac{\pi}{4}\right).
\end{equation} 
Finally we integrate by parts, use property of the Andreev amplitude $a(1)=1$, and drop the remaining part of the integral as it has faster decay with the harmonic index $n$. Thus the supercurrent carried by the continuum of states is given asymptotically by 
the series 
\begin{equation}\label{eq:I-CS-long}
I^{\text{CS}}\approx-\frac{\sqrt{\pi}G\Delta}{e}\left(\frac{E_{\text{Th}}}{\Delta}\right)^{3/2}\sum^{\infty}_{n=1}\frac{1}{n^{3/2}}\cos\left(2n\frac{\Delta}{E_{\text{Th}}}+\frac{\pi}{4}\right)\sin(n\phi),
\end{equation}  
to the leading order in $\Delta/E_{\text{Th}}=L/\xi\gg1$. As compared to the leading part Eq. \eqref{eq:I-sawtooth}, this contribution to the current has faster decay as $\propto 1/L^{3/2}$. Perhaps superficially, but this oscillatory behavior in $I^{\text{CS}}$ is reminiscent to the de Haas-van Alphen (dHvA) effect. In the present context, the spectral edge, $\varepsilon=\Delta$, plays a role similar to that of the Fermi energy $E_{\text{F}}$, while mean-level spacing $E_{\text{Th}}$ is equivalent to the cyclotron energy $\omega_c$ separating Landau levels. Thus, the dHvA magneto-oscillations in the parameter $E_F/\omega_c$ are similar to the current (and force) oscillations in the parameter $\Delta/E_{\text{Th}}$.  

\begin{figure}[t!]
\centering
\includegraphics[width=0.475\linewidth]{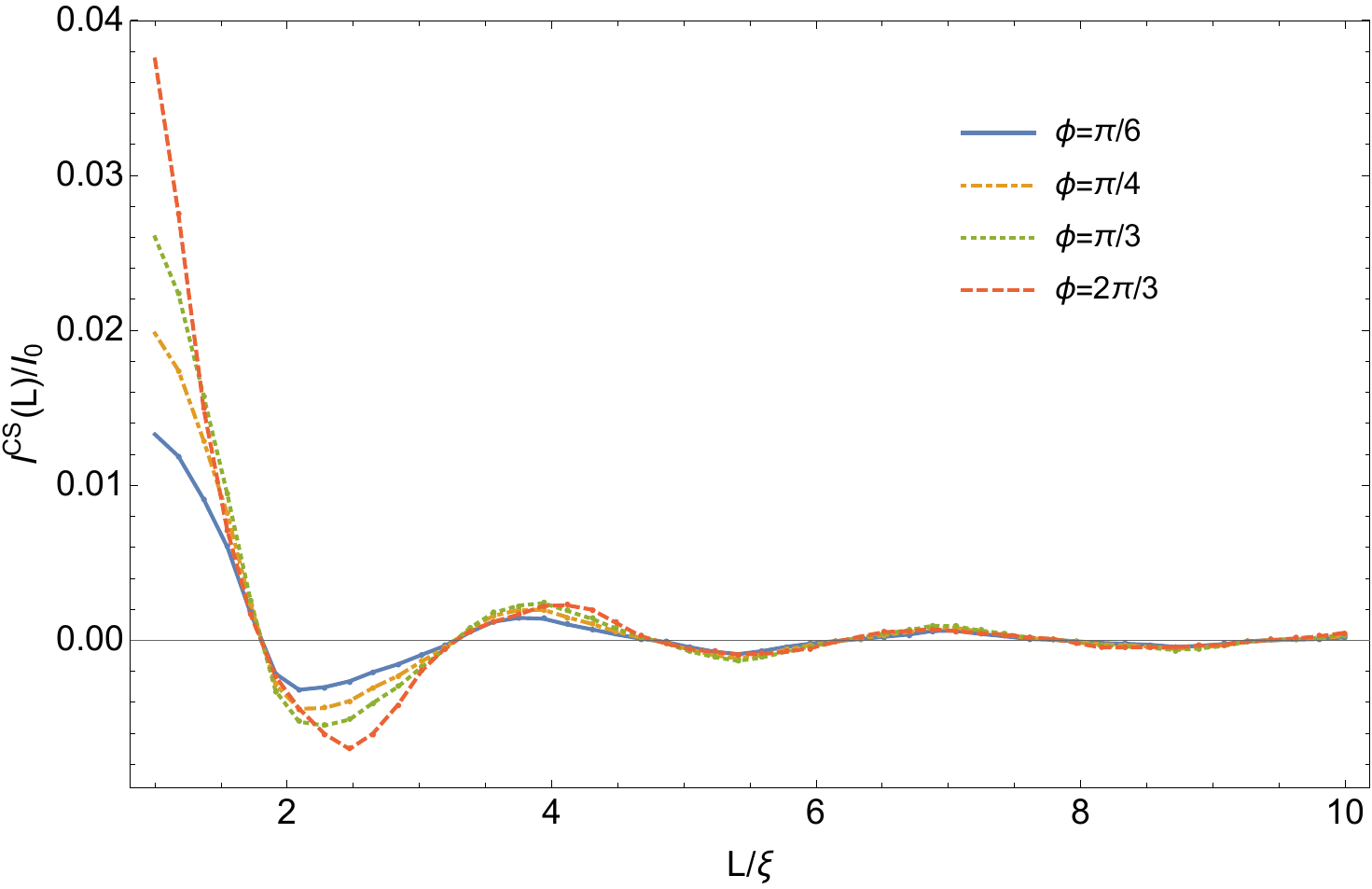}
\includegraphics[width=0.475\linewidth]{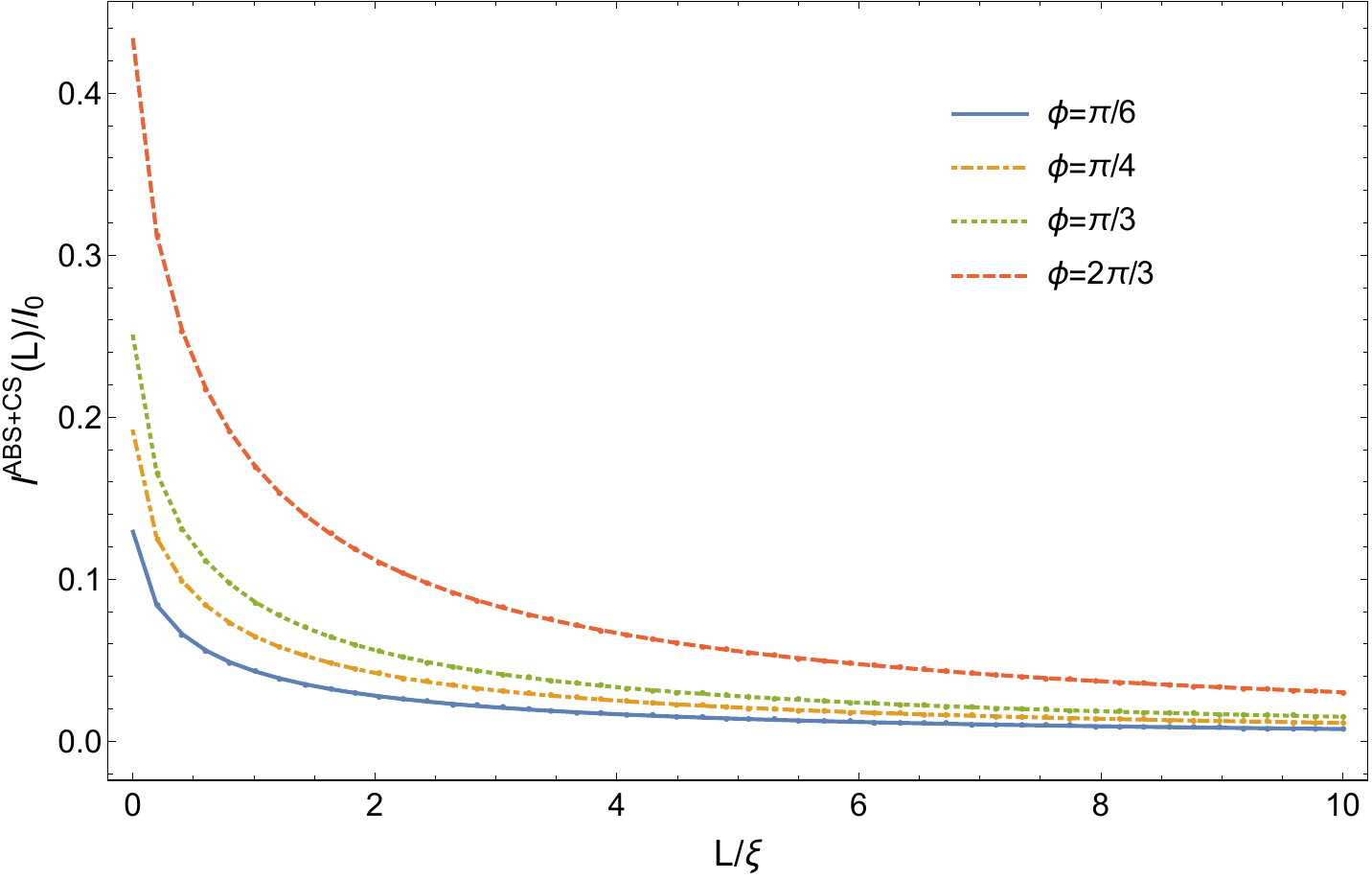}
\caption{The left panel shows the contribution to the Josephson current from the continuum of states, as given by Eq. \eqref{eq:I-square-well}, plotted as a function of junction length in units of the superconducting coherence length. The right panel shows the total current, which includes contributions from the subgap states, resulting in the cancellation of oscillatory terms and a smooth decay of the current for $L\gg\xi$.}\label{fig:I-long-L}
\end{figure}

The approximations leading to Eq. \eqref{eq:In-ABS-long} do not adequately capture bound states with energies close to the gap, $\varepsilon \lesssim \Delta$. In fact, there is a strong tendency for the rapidly oscillatory terms arising from the continuum [Eq. \eqref{eq:I-CS-long}] and the bound states, which are not accounted for in Eq. \eqref{eq:I-sawtooth}, to cancel each other out. This issue, along with the presence of a $1/L^{3/2}$ term in the current-phase relation of long SNS junctions, has been recently discussed (see Refs. \cite{Sonin:2024,Thuneberg:2024}), which also provide a historical overview of the topic. Below, we briefly reiterate the main argument, following the presentation in Refs. \cite{Sonin:2024,Bagwell:1992,Thuneberg:2024}.

Instead of decomposing the Josephson current into Fourier harmonics, one may use a direct integral representation. In 1D geometry of a square-well pair potential model (taking $N_{\text{ch}}=1$ for brevity) it takes the form
\begin{equation}\label{eq:I-square-well}
I=I^{\text{ABS}}+I^{\text{CS}},\quad I^{\text{ABS}}=\frac{2e}{\hbar}\sum_{n,\pm}\frac{d\varepsilon^\pm_{n}}{d\phi}f(\varepsilon^\pm_n),\quad 
I^{\text{CS}}=\frac{2e}{\hbar}\left(\int^{-\Delta}_{-\infty}+\int^{\infty}_{\Delta}\right)\mathcal{I}(\varepsilon,\phi)f(\varepsilon)d\varepsilon
\end{equation} 
The first term, given by a sum, is the contribution from the bound states at $|\varepsilon|<\Delta$ whose energies are found from the Andreev-Kulik quantization condition 
\begin{equation}\label{eq:ABS}
2\arccos(\varepsilon/\Delta)+\frac{2L}{\xi}\frac{\varepsilon}{\Delta}\pm\phi=2\pi n, \quad n\in\mathbb{Z}. 
\end{equation}
The second term, is the sum of two integrals over the continuum of states below and above the gap with the spectral weight given by 
\begin{equation}
\mathcal{I}(\varepsilon,\phi)=\frac{(2\Delta^2/\pi)|\varepsilon|\sqrt{\varepsilon^2-\Delta^2}\sin\frac{2\varepsilon}{E_{\text{Th}}}\sin\phi}{\left[(\Delta^2-2\varepsilon^2)\cos\frac{2\varepsilon}{E_{\text{Th}}}+\Delta^2\cos\phi\right]^2+
4\varepsilon^2(\varepsilon^2-\Delta^2)\sin^2\frac{2\varepsilon}{E_{\text{Th}}}}.
\end{equation}  
In each of these expressions, the thermal occupation in equilibrium is governed by the Fermi function $f(\varepsilon)=1/(e^{\varepsilon/T}+1)$, however Eq. \eqref{eq:I-square-well} is applicable for arbitrary nonequilibrium distribution $f(\varepsilon)$. 

From the antisymmetry of $\mathcal{I}(\varepsilon,\phi)$ with respect to the interchange $\varepsilon\to-\varepsilon$, one can combine both integrals to have a single integration going over the positive energies only, with the thermal factor $f(-\varepsilon)-f(\varepsilon)=2f(\varepsilon)-1$. In this representation it is obvious that $I^{\text{CS}}$ is a periodic function in $L/\xi$, when we make energy integration to be dimensionless in $x=\varepsilon/\Delta$, due to the presence of oscillatory functions $\sin(2xL/\xi)$ and $\cos(2xL/\xi)$ under the integral. The result of the numerical integration of $I^{\text{CS}}$ is shown in Fig. \ref{fig:I-long-L} for several values of $\phi$. 

\begin{figure}[t!]
\centering
\includegraphics[width=0.475\linewidth]{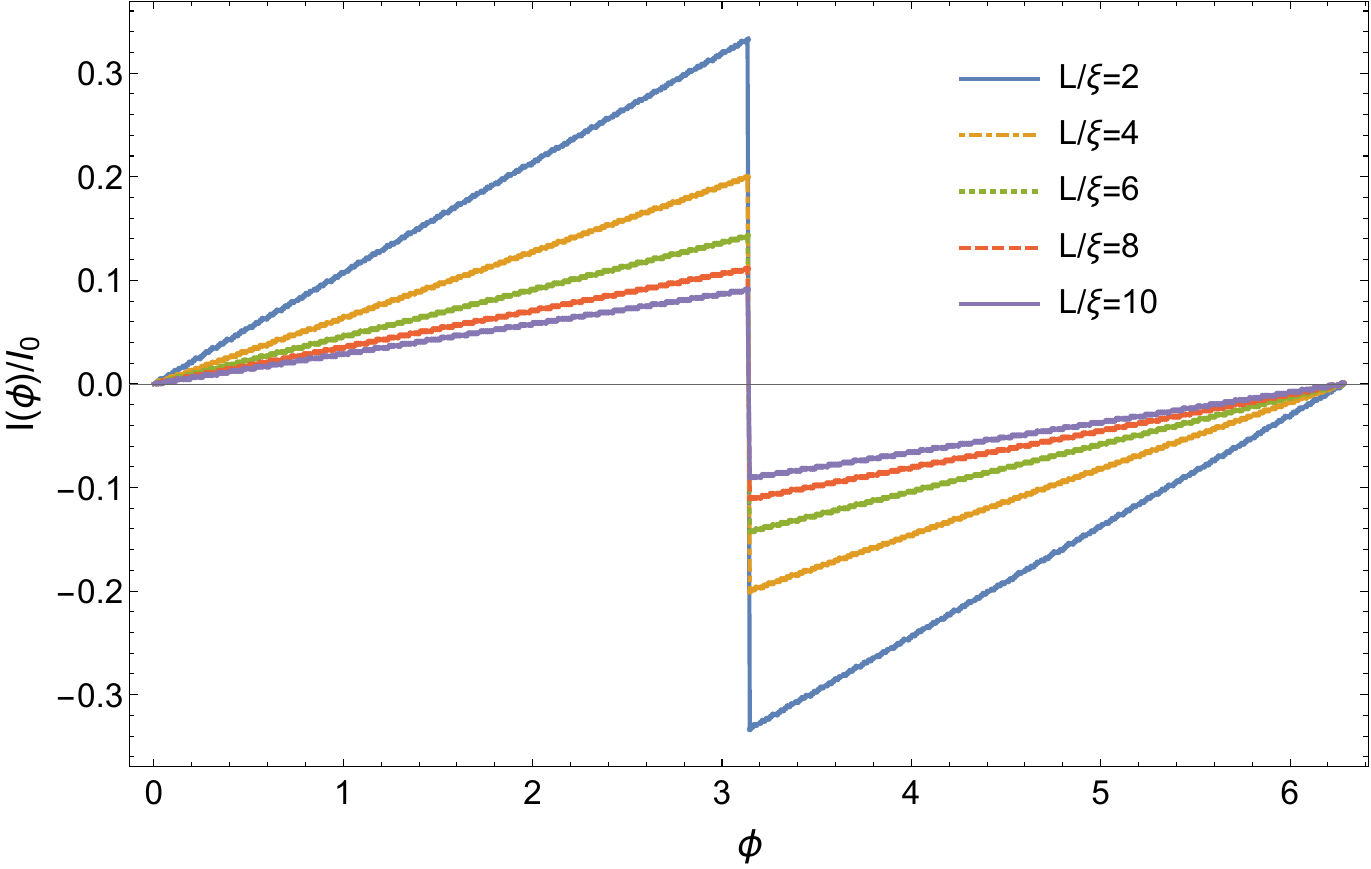}
\includegraphics[width=0.475\linewidth]{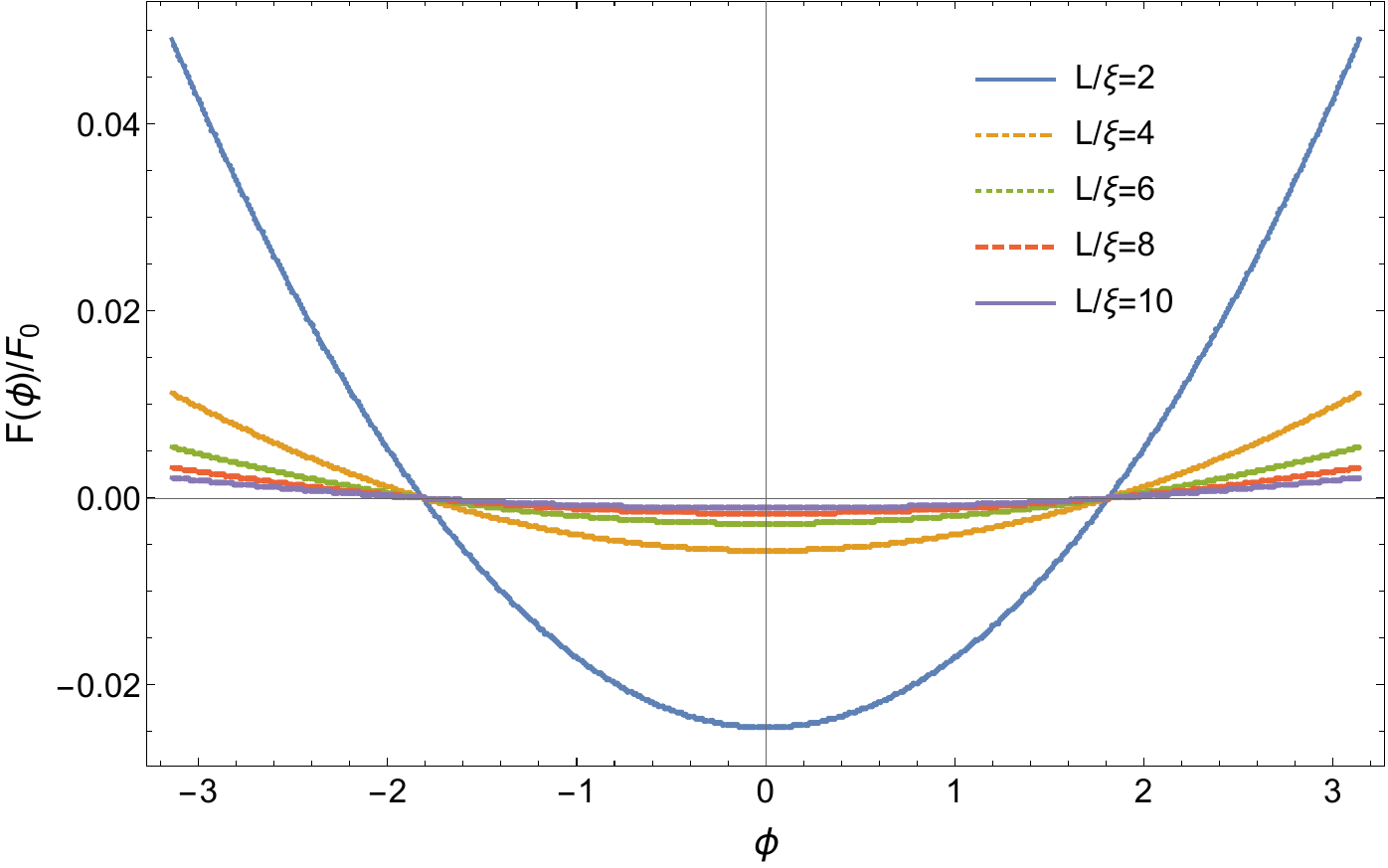}
\caption{The left panel shows the total current-phase relation, including both the bound state and continuum contributions, plotted for various junction lengths as indicated in the plot legends. The right panel shows the phase dependence of the conjugate force, which includes subleading correction terms in $\xi/L\ll1$, up to quadratic order.}\label{fig:I-long-phi}
\end{figure}

On the other hand, one can convert the energy integral from real energies to the Matsubara axis $\varepsilon=i\varepsilon_m$. This allows a single compact expression for the current that includes both bound states and continuum contributions \cite{Ishii:1970,Svidzinsky,Thuneberg:2024}
\begin{equation}\label{eq:I-Matsubara}
I=\frac{2eT}{\hbar}\sum^{\infty}_{m=-\infty}\frac{\Delta^2\sin\phi}{(2\varepsilon^2_m+\Delta^2)\cosh\frac{2\varepsilon_m}{E_{\text{Th}}}+2\varepsilon_m\sqrt{\varepsilon^2_m+\Delta^2}\sinh\frac{2\varepsilon_m}{E_{\text{Th}}}+\Delta^2\cos\phi}. 
\end{equation}    
In the limit of zero temperature, the sum converts back into the integral over the real axis, $T\sum_m(\ldots)\to \int\frac{d\varepsilon}{2\pi}(\ldots)$. As a result, the total current, the sum of contributions of bound states and the continuum, is found to be a monotonically decaying function with increasing length of the junction. This is comparatively shown in Fig. \ref{fig:I-long-L}. Equation \eqref{eq:I-square-well} is more physically transparent, while Eq. \eqref{eq:I-Matsubara} is more efficient for the numerical implementation. For completeness, in Fig. \ref{fig:I-long-phi} we show the current-phase relation and the phase dependence of the cohesive force in the long junction limit.   

%############################ 
\subsection{Impure junctions}
%############################ 

In this section, we examine the effect of impurities on the Casimir effect in a Josephson junction. To reach the main conclusion, it is sufficient to consider the simplest possible model: a single scatterer in the junction. Following the classical work on the subject (Ref. \cite{Bagwell:1992}), we model a point-like impurity in the SNS junction using a delta-function potential, $V(x)=V_0\delta(x-a)$, where $0<a<L$. The transmission coefficient for scattering on this impurity potential is given by
\begin{equation}
\mathcal{T}=1-\mathcal{R}=\frac{1}{1+(mV_0/\hbar^2k_{\text{F}})^2}.
\end{equation}   
In the presence of impurity, the quantization of Andreev-Kulik energy levels is modified. Instead of Eq. \eqref{eq:ABS} one finds 
\begin{equation}\label{eq:ABS}
2\arccos(\varepsilon/\Delta)+\frac{2L}{\xi}\frac{\varepsilon}{\Delta}\pm\varphi=2\pi n, \quad n\in\mathbb{Z}. 
\end{equation}
where the effective phase $\varphi$ is determined from the equation
\begin{equation}
\cos\varphi=\mathcal{T}\cos\phi+\mathcal{R}\cos\left(2\frac{L-2a}{\xi}\frac{\varepsilon}{\Delta}\right).
\end{equation}
In the limit of a point contact, $L/\xi\to0$, at finite transmission, $\mathcal{T}<1$, the current is independent of $L$, therefore, the conjugated force vanishes to the leading oder. 
At finite $L/\xi\ll1$ the leading order correction originates from the continuum. The corresponding part of the supercurrent can be expressed as follows 
\begin{equation}\label{eq:I-CS-impurity}
I^{\text{CS}}=\frac{ge}{2\pi\hbar}\mathcal{T}\int^{\infty}_{\Delta}[u^2(\varepsilon)-v^2(\varepsilon)][\mathcal{D}^{-1}(\varepsilon,-\varphi)-\mathcal{D}^{-1}(\varepsilon,\varphi)]
\left(\frac{\sin\phi}{\sin\varphi}\right)[2f(\varepsilon)-1]d\varepsilon,
\end{equation}    
thus generalizing the corresponding term in Eq. \eqref{eq:I-square-well}, where the coherence factors are given by 
\begin{equation}
u^2(\varepsilon)=\frac{1}{2}\left[1+\frac{\sqrt{\varepsilon^2-\Delta^2}}{\varepsilon}\right],\quad v^2(\varepsilon)=\frac{1}{2}\left[1-\frac{\sqrt{\varepsilon^2-\Delta^2}}{\varepsilon}\right],
\end{equation}
and the spectral factor is given by 
\begin{equation}
\mathcal{D}(\varepsilon,\varphi)=u^4(\varepsilon)+v^4(\varepsilon)-2u^2(\varepsilon)v^2(\varepsilon)\cos\left[2\frac{\varepsilon}{\Delta}\frac{L}{\xi}+\varphi\right].
\end{equation}
An inspection of Eq. \eqref{eq:I-CS-impurity} reveals that the most significant range of energies that predominantly determines the length dependence of $I^{\text{CS}}$, and thus the force, is $\Delta\ll\varepsilon\ll E_{\text{Th}}$.  
In this domain 
\begin{equation}
\mathcal{D}(\varepsilon,\pm\varphi)\approx 1-\frac{\Delta^2}{2\varepsilon^2}\left[\cos\varphi\mp2\frac{\varepsilon}{\Delta}\frac{L}{\xi}\sin\varphi\right],
\end{equation}
so that to the leading order in logarithmic accuracy 
\begin{equation}
I^{\text{CS}}\approx-\frac{ge}{2\pi\hbar}2\mathcal{T}\sin\phi\frac{L}{\xi}
\int^{\sim E_{\text{Th}}}_{\Delta}\frac{\Delta\sqrt{\varepsilon^2-\Delta^2}}{\varepsilon^2}d\varepsilon\simeq-\frac{2e}{\hbar}\mathcal{T}\frac{g\Delta}{2\pi}\frac{L}{\xi}\ln\frac{\xi}{L}\sin\phi.
\end{equation}
Finally, for the force, with the accuracy up to $\phi$-independent terms, one finds
\begin{equation}
F\approx\mathcal{T}\frac{g\Delta}{\pi\xi}\ln\frac{\xi}{L}\sin^2(\phi/2).
\end{equation}  
This result should be contrasted to the earlier finding of Ref. \cite{Krive:2004a} where it was found that the tunnel barrier reduces the $\phi$ dependence of the force by a factor of $\sqrt{\mathcal{T}}$. This conclusion pertains only to the consideration of the force arising from the bound states. In contrast, our conclusions agrees with Ref. \cite{Beenakker:2023} showing that the force is more sensitive to the presence of a tunnel barrier decreasing as $\propto \mathcal{T}$ rather than $\sqrt{\mathcal{T}}$.  

%#########################################################################
%#########################################################################
%#########################################################################
%#########################################################################
%#########################################################################

\section{Summary discussion and estimates}\label{sec:Summary}

In this work, we explored the peculiarities of the Casimir effect in Josephson junctions and calculated the corresponding force in the limits of short and long ballistic multichannel contacts, as well as impure constrictions. 

In short junctions, the force is dominated by contributions from the continuum of states above the superconducting gap. This contrasts with the behavior of the Josephson current, which is predominantly determined by the sub-gap states, receiving only small corrections from the state above the gap. The Andreev-Casimir force in this limit saturates to an almost universal magnitude, per transport channel, determined solely by the properties of the superconductor -- namely its gap and coherence length. The characteristic scale of the force is on the order of $\sim N_{\text{ch}}\Delta/\xi$. 

In junctions longer than the coherence length, the contribution to the force from the continuum exhibits pronounced oscillations as a function of the junction length and the superconducting phase difference across the junction. This contribution is canceled by the corresponding oscillatory component arising from the bound states. As a result, the force exhibits only a smooth power-law decay, scaling as $F\propto 1/L^2$ to leading order.

It should be expected that scattering in the contact would strongly influence the magnitude of the force. We find that at finite transmission $\mathcal{T}<1$ the force is indeed reduced. In a model of a single scatterer in the junction it scales linearly with $\mathcal{T}$. 

In a diffusive superconducting point contact, $l\ll L\ll \xi'$, where $l$ is the mean free path and $\xi'=\sqrt{\xi l}$ is the coherence length in the disordered limit, the critical current amplitude saturates to the value $\sim G\Delta/e$, with the conductance $G\sim (e^2/\hbar)N_{\text{ch}}(l/L)$. Therefore, in this case the corresponding force is dominated by the $L$ dependence of the conductance, one estimates $F\sim N_{\text{ch}}\Delta l/L^2$. For longer junctions, $l\ll\xi'\ll L$, the amplitude of the current is determined by the mini-gap induced by a proximity effect in the normal region of the junction, $\sim G E_{\text{g}}/e$ \cite{Schon:2001}, which is of the order of diffusive Thouless energy $E_{\text{g}}\sim D/L^2$, where $D=v_Fl$ is the diffusion coefficient in the wire. As a result, the force can be estimated $F\sim N_{\text{ch}}E_{\text{g}}l/L^2\propto 1/L^4$, which is negligibly small.    

Taking optimistic values for the gap $\Delta\sim 10$ K and coherence length $\xi\sim 10$ nm of a conventional superconductor, the estimated force is at best $\Delta/\xi\lesssim 0.1$ pN (these numbers are close to those observed in superconducting niobium nanofilms \cite{Pinto:2018}). Due to this strong limitation, a very large number of channels is required to boost the force magnitude closer to measurable values. For example, in a junction with  $N_{\text{ch}}\sim 10^3$, and considering the logarithmic enhancement factor in the force [Eq. \eqref{eq:F-CS-short}] which can add another factor of 10, one might hope to achieve a force magnitude of $F\sim 1$ nN. 
It has been suggested that modern low-temperature STM and AFM techniques with superconducting tips could be used to measure such forces. While the measurement of the intrinsic superconducting properties 
via Josephson scanning microscopy has been successfully implemented, see e.g. \cite{Dynes:2001,Sacks:2006,Dynes:2009,Yazdani:2016,Senkpiel:2020,Liu:2021}, measuring the corresponding force remains a significant challenge, but a substantial progress has been achieved in that direction \cite{Wilson:2011,Hakonen:2013}. It should be noted that the estimated magnitude of the Andreev-Casimir force is well within the precision measurement capabilities of modern apparatus. For instance, AFM can detect forces as weak as $\sim10$ pN when operated in ultrahigh vacuum at a temperature of $\sim 5$ K \cite{Ternes:2008}.

In closing, we note that electron interactions may strongly influence the magnitude of the Andreev-Casimir force. This is, however, a much more sophisticated problem to consider. There are studies of the Josephson effect in Luttinger liquids  \cite{Maslov:1996}, where interactions can be accounted for beyond perturbation theory via bosonization, with even exact results for the supercurrent obtained analytically through the thermodynamic Bethe ansatz for a solvable two-boundary version of the sine-Gordon model \cite{Caux:2002}. In the context of the Luttinger liquids model, this latter case corresponds to attractive interactions and is not immediately applicable to electron transport in quantum wires. Other interesting examples of Casimir forces in interacting systems include quantum impurity problems \cite{Bulgac:2005,Fuchs:2007,Zhabinskaya:2008,Schecter:2014}.

\bmhead{Acknowledgements}

I am grateful to Carlo Beenakker for sharing the results of Ref. \cite{Beenakker:2023} prior to publication, which drew my attention to this interesting problem, and for the constructive communication that followed. I also thank Anton Andreev and Alex Kamenev for useful discussions, and Victor Brar and Roman Kuzmin for pointing out relevant experimental literature. Finally, I am grateful to Elio K\"onig for reading and providing feedback on the paper. 

This work was financially supported by the National Science Foundation, Quantum Leap Challenge Institute for Hybrid Quantum Architectures and Networks Grant No. OMA-2016136 and H. I. Romnes Faculty Fellowship provided by the University of Wisconsin-Madison Office of the Vice Chancellor for Research and Graduate Education with funding from the Wisconsin Alumni Research Foundation.

\bibliography{biblio}
\end{document}